\def\year{2026}\relax

\documentclass[letterpaper]{article} 
\usepackage{aaai25}  
\usepackage{times}  
\usepackage{helvet}  
\usepackage{courier}  
\usepackage[hyphens]{url}  
\usepackage{graphicx} 
\usepackage{natbib}  
\usepackage{caption} 
\DeclareCaptionStyle{ruled}{labelfont=normalfont,labelsep=colon,strut=off} 
\usepackage{algorithm}
\usepackage{algorithmic}

\usepackage{xcolor}

\usepackage{newfloat}
\usepackage{listings}
\floatstyle{ruled}
\newfloat{listing}{tb}{lst}{}
\floatname{listing}{Listing}
\nocopyright
\usepackage{multirow}
\usepackage{tabularx}
\usepackage{amssymb}
\usepackage{amsmath}
\usepackage{nameref}

\usepackage[detect-all]{siunitx}
\usepackage{eurosym}
\DeclareSIUnit{\sieuro}{\mbox{\euro}}

\usepackage{xspace}
\newcommand\ie{i.\,e.\xspace} 
\newcommand\eg{e.\,g.\xspace}

\def\sym#1{\ifmmode^{#1}\else\(^{#1}\)\fi}

\graphicspath{{./img/}{docs/685c853d7b309ec44e0c8ac1/img/}}

\usepackage[toc,page]{appendix}

\usepackage[en-US]{datetime2}
\usepackage[shortcuts]{extdash}

\usepackage{enumitem}

\title{Supply vs. Demand in Community-Based Fact-Checking on Social Media}
\author{
    Moritz Pilarski\textsuperscript{\rm 1},
    Nicolas Pr{\"o}llochs\textsuperscript{\rm 1}}
\affiliations{
    \textsuperscript{\rm 1}JLU Giessen, Germany\\
    moritz.pilarski@wi.jlug.de, nicolas.proellochs@wi.jlug.de
}

\begin{document}

\maketitle

\begin{abstract}
    Fact\-/checking ecosystems on social media depend on the interplay between what users want checked and what contributors are willing to supply. Prior research has largely examined these forces in isolation, yet it remains unclear to what extent supply meets demand. We address this gap with an empirical analysis of a unique dataset of 1.1 million fact\-/checks and fact\-/checking requests from X's Community Notes platform between June 2024 and May 2025. We find that requests disproportionately target highly visible posts -- those with more views and engagement and authored by influential accounts -- whereas fact\-/checks are distributed more broadly across languages, sentiments, and topics. Using a quasi\-/experimental survival analysis, we further estimate the effect of displaying requests on subsequent note creation. Results show that requests significantly accelerate contributions from Top Writers. Altogether, our findings highlight a gap between the content that attracts requests for fact\-/checking and the content that ultimately receives fact\-/checks, while showing that user requests can steer contributors toward greater alignment. These insights carry important implications for platform governance and future research on online misinformation.
\end{abstract}

\section{Introduction}


The spread of misinformation on social media poses serious risks for modern societies \cite{Bar.2023,Ecker.2024,Altay.2025,WEF.2024}. In response, social media platforms have invested in fact\-/checking initiatives that range from collaborations with professional fact\-/checkers to community\-/based approaches that enlist users themselves \cite{Prollochs.2022a,Allen.2021}. For the latter, a prominent example is Community Notes on X (formerly Twitter), where contributors collaboratively identify and annotate misleading content \cite{Prollochs.2022a,Twitter.2021}. Compared to professional fact\-/checking, community\-/based systems promise greater scalability, producing a larger number of fact\-/checks at higher speed \cite{Pennycook.2019, Chuai.2024b}. 


Yet, a key requirement for the effectiveness of fact\-/checking systems is the balance between \emph{supply} (\ie, the fact\-/checks contributors provide) and \emph{demand} \cite[\ie, the content users want to see fact\-/checked;][]{Guess.2020,Graham.2025}. Because fact\-/checkers, whether professional or community\-/based, have limited time and resources, it is unrealistic to verify all (or even most) content \cite{Graham.2025,Guo.2022}. Their efforts must therefore be allocated carefully across content: when supply is misaligned with demand, high\-/priority posts may receive disproportionately little attention, while lower\-/priority content may attract relatively more fact\-/checking effort than users perceive as warranted. 
However, despite the centrality of this balance, systematic evidence on how fact\-/checking demand and supply align in real\-/world social media platforms is largely missing.


Prior research has typically examined either demand or supply in isolation, rather than how these forces interact. Work on the \emph{supply} side has primarily analyzed what types of content fact\-/checkers choose to verify. Studies find, for example, that fact\-/checkers tend to prioritize highly viral claims \cite{Pilarski.2024,Bond.2023} and that right\-/leaning misinformation is disproportionately targeted for verification \cite{Chuai.2025b,Renault.2025}. Research on the \emph{demand} side is more limited and mostly comprises surveys measuring public support for fact\-/checking \cite{Rich.2020,Martel.2024,Horne.2025,Reuters.2025}. For instance, a 2025 survey reports that \SI{73}{\percent} of respondents in the United States are concerned about distinguishing real from fake news online -- a sharp increase from 2023 -- and that social media is perceived as the primary source of misinformation \cite{Reuters.2025}. Large majorities of the population believe that platforms should reduce the spread of harmful misinformation, express support for warning labels on misleading posts, and endorse platform\-/level interventions \cite{Martel.2024,Horne.2025}. Yet despite these indications of high public demand, we lack systematic evidence on whether the supply of fact\-/checks aligns with what users consider important on real\-/world social media platforms. This gap motivates our study, which provides the first large\-/scale empirical analysis of fact\-/checking contributions and requests on the social media platform X (formerly Twitter).


\textbf{Research goal:} In this study, we provide the first large\-/scale empirical analysis of supply and demand in fact\-/checking on social media. To this end, we leverage the recently introduced requests feature on X's Community Notes platform, which enables users to directly request fact\-/checks on specific posts. This feature offers a unique, large\-/scale opportunity to measure demand empirically and to examine how it aligns with the supply of fact\-/checks.
Specifically, we address the following research questions:

\begin{enumerate}[
    label=\textbf{RQ\,\arabic*:},
    align=left,
    left=0pt,
    labelwidth=*,
    topsep=0.5\baselineskip,
    midpenalty=10000 
]
\item \emph{To what extent does fact\-/checking supply align with user demand?}
\item \emph{How do fact\-/checking requests influence contributors' fact\-/checking activity?}
\end{enumerate}

\textbf{Data \& analysis:} We collected a unique dataset of 1.1 million fact\-/checks and fact\-/checking requests from X's Community Notes platform between June 2024 and May 2025, \ie, for an observation period of one year. Based on this data, we compare the characteristics of posts targeted by requests (demand) and those that ultimately receive fact\-/checks (supply). Here, we focus on a wide variety of post and author characteristics, including engagement, social influence, sentiment, and topics. Furthermore, using a quasi\-/experimental design, we estimate the effect of displaying user requests on subsequent fact\-/check creation. The data collection and the analysis follow common standards for ethical research \cite{Rivers.2014}.

\textbf{Contributions:} Our study is the first large\-/scale study comparing supply vs. demand in fact\-/checking on social media. We find that fact\-/checking requests disproportionately target highly visible posts (\ie, those with more views and engagement and authored by influential accounts), whereas fact\-/checks are distributed more broadly across languages, sentiments, and topics. We further show that displaying fact\-/checking requests significantly accelerates fact\-/checking contributions. Taken together, these findings indicate a gap between the content users want most fact\-/checked and the content that ultimately receives fact\-/checks, while also demonstrating that request mechanisms can steer contributors toward greater alignment. Our work has important implications for platform design and future research on misinformation.

\section{Background}

\textbf{Misinformation on social media:} 
Social media platforms have become central gateways to information access and public debate \cite{VanBavel.2024, Pew.2024b}. Because anyone can post and share content, social media enables information to spread widely and quickly \cite{Lazer.2018, Shore.2018, Kim.2019}. Unlike traditional media, however, there is little expert oversight or editorial control, leaving platforms particularly vulnerable to the circulation of misinformation \cite{Shao.2016, Vosoughi.2018}. Empirical work has shown that false information often spreads faster and more broadly than true information \cite{Vosoughi.2018, Solovev.2022b, Prollochs.2021a, Prollochs.2022b}. This is problematic as exposure to misinformation has been linked to harmful outcomes, including distorted political perceptions during elections \cite{Allcott.2017, Bakshy.2015, McCabe.2024} and risky behaviors during public health crises \cite{Gallotti.2020, Pennycook.2020b, Solovev.2022b}. The challenge is further amplified by advances in AI, which enable the creation of misinforming content at unprecedented speed and scale \cite{Feuerriegel.2023,Drolsbach.2025}.

Research has identified several drivers of misinformation diffusion. False content is often designed to deceive, making it difficult for users to identify inaccuracies \cite{Wu.2019}. Furthermore, many users rarely check the veracity of what they encounter online \cite{Geeng.2020, Vo.2018}, with research suggesting that limited cognitive reflection impedes critical evaluation \cite{Moravec.2019, Pennycook.2019b, Pennycook.2021}. Social networks also exhibit strong homophily, as people tend to connect with those who share similar worldviews \cite{McPherson.2001}. This reinforces echo chambers, where users are disproportionately exposed to like\-/minded perspectives \cite{Barbera.2015}. Together, these dynamics create fertile ground for misinformation to spread and persist, underscoring the need for effective countermeasures. 

\textbf{Fact\-/checking:}
Effectively countering misinformation requires fact\-/checking strategies that are both accurate and scalable. Existing approaches can be roughly grouped into three categories. First, third\-/party fact\-/checking organizations such as \emph{politifact.com} and \emph{snopes.com} provide highly accurate assessments, but they cannot keep pace with the sheer volume of misinformation and are often distrusted by users who perceive them as biased \cite{Wu.2019, Micallef.2022, Pennycook.2019, Poynter.2019, Drolsbach.2024}. Second, automated machine learning approaches can operate at scale but typically sacrifice accuracy \cite{Ma.2016, Wu.2019}. Third, community\-/based fact\-/checking systems delegate fact\-/checking to platform users themselves \cite{Allen.2021, Bhuiyan.2020, Pennycook.2019, Prollochs.2022a, Drolsbach.2023, Drolsbach.2023b}. By aggregating the judgments of diverse users, these systems draw on the ``wisdom of crowds'' \cite{Allen.2021, Bhuiyan.2020, Pennycook.2019, Prollochs.2022a}. Research shows that the accuracy of such crowd assessments can rival expert judgments, even when generated by relatively small groups \cite{Bhuiyan.2020, Epstein.2020, Resnick.2021}.

\begin{figure*}[!t]
    \centering
    \includegraphics[width=\linewidth, keepaspectratio]{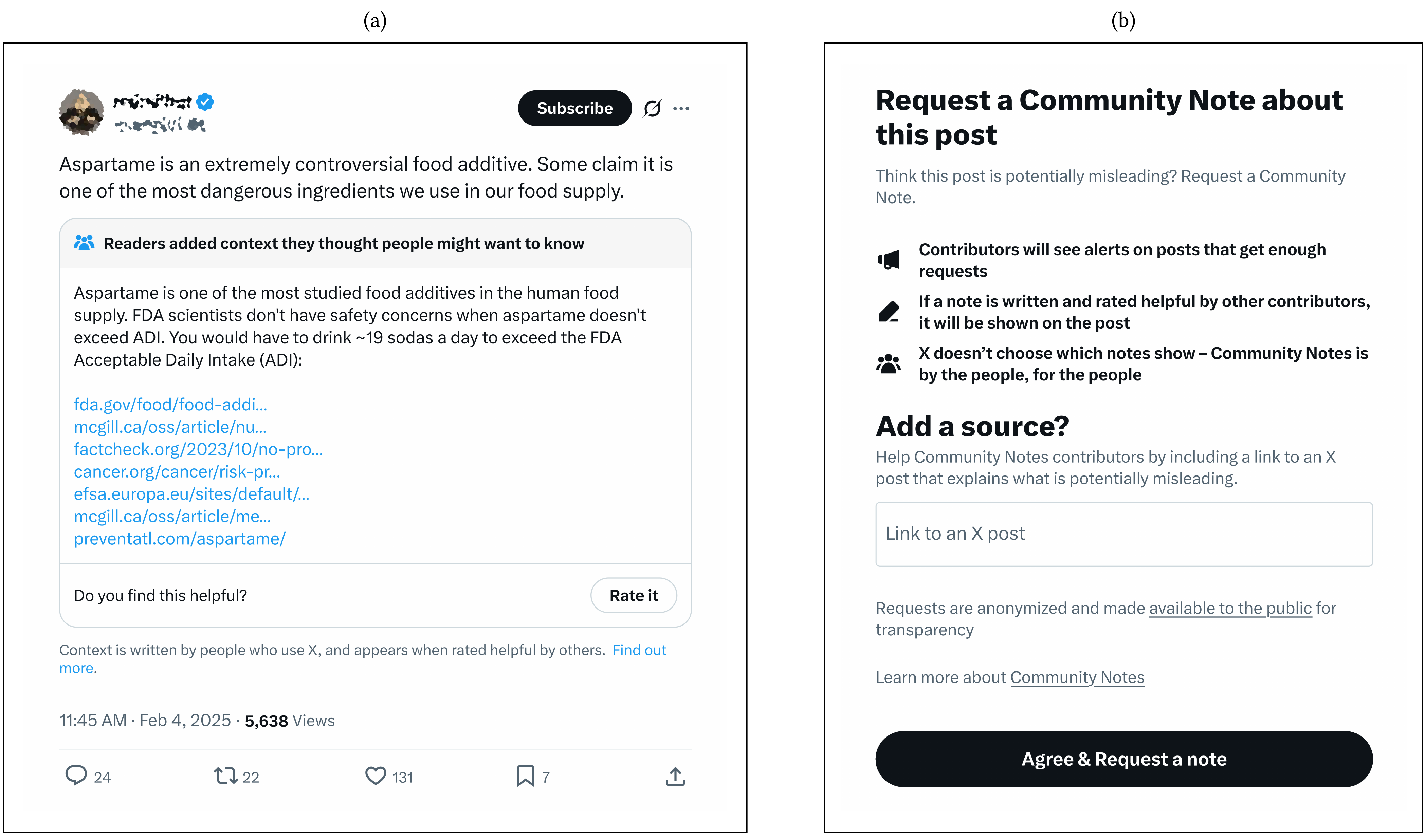}
    \caption{Screenshots of (a) a post with a displayed Community Note and of (b) the Community Notes request dialog.}
    \label{fig:screenshot}
\end{figure*}

\textbf{Community Notes:}
A prominent implementation of a community\-/based approach to fact\-/checking is Community Notes on X (formerly Twitter), launched globally in December 2022. Different from earlier crowd\-/based fact\-/checking initiatives, Community Notes allows users to identify misinformation directly on the platform. In Community Notes, registered contributors can attach short contextual notes to posts they perceive as either misleading or not misleading (see example in Fig.~\ref{fig:screenshot}a), explaining why the content may be inaccurate \cite[\eg, by referencing relevant sources;][]{Prollochs.2022a,Twitter.2021,CNsFAQ.2024, Solovev.2025}. 
Other contributors then see an indicator that a post has a note and are prompted to rate its helpfulness. Notes become publicly visible only when they label a post as misleading and contributors with historically diverse rating profiles agree that the note is helpful -- a design intended to counteract unilateral ideological influence and promote cross\-/perspective validity \cite{Twitter.2021,CNsFAQ.2024}.

Compared to professional fact\-/checking, Community Notes substantially expand coverage by enabling a larger volume of content to be annotated at higher speed \cite{Pennycook.2019,Chuai.2024}. Although concerns about political bias persist \cite{Allen.2022,Prollochs.2022a}, studies consistently find that users perceive Community Notes as informative, helpful, and trustworthy \cite{Prollochs.2022a,Drolsbach.2023b,Solovev.2025}. Experimental evidence further shows that Community Notes reduce users' belief in false claims and their intentions to share misleading posts. Complementing this, recent field studies demonstrate that displaying notes beneath posts can causally and substantially curb the spread of misinformation by reducing subsequent reposting; an effect that holds across both sides of the political spectrum \cite{Chuai.2024,Slaughter.2025}. Altogether, community\-/based fact\-/checking systems such as X's Community Notes program may have the potential to address many of the drawbacks of traditional approaches to fact\-/checking on social media.

\textbf{Supply and demand in fact\-/checking:} 
Prior research has examined the \emph{supply} of fact\-/checks, \ie, which types of content are selected for verification and by whom \cite{Pilarski.2024,Renault.2025,Chuai.2025b}. Work in this area shows, for example, that right\-/leaning misinformation is more frequently targeted by both professional and community fact\-/checkers \cite{Renault.2025,Chuai.2025b}. On professional fact\-/checking sites, mentions of political elites in fact\-/checked false statements peak in the months preceding elections, and false statements are generally more likely to reference political elites than true ones \cite{Chuai.2025b}. On X, fact\-/checking activity is heavily concentrated in political domains and disproportionately directed at high\-/engagement posts and users \cite{Vosoughi.2018,Pilarski.2024}. Content with emotional or negative tone and impoliteness is also more likely to attract corrective responses \cite{Ma.2023}. On Reddit, commenters post fact\-/check links sooner on false stories than on true ones \cite{Bond.2023}. Overall, existing research suggests that fact\-/checking supply is not random; instead, verification efforts systematically concentrate on specific types of content, users, and contexts.

In contrast, research on the \emph{demand} for fact\-/checking is relatively scant. Most existing studies rely on surveys that assess general public attitudes, consistently finding broad support for fact\-/checking interventions across the political spectrum \cite{Rich.2020,Horne.2025}. A recent global survey reported that \SI{58}{\percent} of respondents are concerned about distinguishing between real and fake news online, with concern rising to \SI{73}{\percent} in the United States. 
Social media platforms were most frequently cited as key channels for the spread of misleading information \cite{Reuters.2025}. Additional studies show that large majorities believe platforms should take steps to reduce harmful misinformation, support the use of warning labels on misleading posts, and endorse broader platform\-/level interventions \cite{Martel.2024,Horne.2025}. Another stream of research has tested interventions to stimulate demand, such as appeals to civic duty \cite{Chopra.2022,Graham.2025}. Evidence from messaging environments also offers indirect signals of demand: during India's 2019 election, for example, a WhatsApp tipline received substantial volumes of forwarded rumors, and much of the viral misinformation circulating in public groups had already been submitted by users for verification \cite{Kazemi.2022}. However, until the introduction of the notes request feature on X, major social media platforms offered no mechanism for users to directly request fact\-/checks, leaving the real\-/world demand for fact\-/checks largely unobservable. 

\textbf{Our contribution:} Prior research has examined fact\-/checking demand or supply in isolation, but not how these forces interact. Our study addresses this gap by conducting a large\-/scale empirical analysis of fact\-/checking contributions and requests on the social media platform X (formerly Twitter). To the best of our knowledge, our study is the first large\-/scale study comparing supply vs. demand in fact\-/checking on social media.

\section{Data \& Methods}

\subsection{Data Sources}

We study fact\-/checking on X's Community Notes platform by comparing two distinct data sources: (i) \emph{fact\-/check supply}, captured by contributor\-/written notes, and (ii) \emph{fact\-/check demand}, reflected in user\-/submitted requests. Our dataset combines both sources and links them back to the underlying posts.

\textbf{Notes:} On X, users can register as Community Notes contributors. Contributors may write fact\-/checks/notes on posts adjudged either ``misleading'' or ``not misleading''. Other contributors can see that a post has a proposed Community Note and can rate the note as \emph{Helpful} or \emph{Not Helpful}. If enough contributors from sufficiently diverse points of view rate a note as helpful, the note becomes publicly visible on the post (i.e., visible to all users, not only registered contributors; see Fig.~\ref{fig:screenshot}a). The diversity of viewpoints is inferred from the historical rating behavior of the raters \cite{CoNo.Docs.Diversity}.

\textbf{Requests:} 
On \DTMdate{2024-06-24}, X introduced a feature that allows all users -- not only registered contributors -- to request Community Notes \citep{CoNo.Docs.Request}. Requests indicate that users believe a post may be misleading (see Fig.~\ref{fig:screenshot}b). When a post younger than \num{24} hours meets the request display threshold  \citep[defined by both the number of requests and their rate relative to views;][]{CoNo.Docs.Request}, the platform shows a callout beneath the post and lists the post in the ``Note Requests'' feed. Visibility of these displays is restricted to contributors with \emph{Top Writer} status, determined by a contributor's history of writing helpful notes \citep{CoNo.Code.TopWriter,CoNo.Docs.TopWriter}. During the observation period of this study, the request feature remained in a pilot phase. The exact rules have since slightly evolved, but retain the same core idea: to surface posts where many users request fact\-/checking.

\subsection{Data Collection}

We use the Community Notes database dump from \DTMdate{2025-05-21}, which includes all notes and requests created up to \DTMdate{2025-05-19} \citep{CoNo.Docs.Download}. Our analysis focuses on notes in which contributors adjudged the target post to be \emph{misleading}, along with all note requests submitted between \DTMdate{2024-06-24} (\ie, when the requests feature was introduced) and \DTMdate{2025-05-19}. This raw dataset contains \num{672732} notes and \num{5898267} requests.

For these notes and requests, we then collected data on the corresponding source posts (\ie, the posts targeted for fact\-/checking or requested by users). 
We first retrieved all posts associated with the \num{672732} note events via the X REST API, excluding cases where the post was unavailable\footnote{Posts are unavailable via the X API if they are deleted, protected, or from suspended accounts.}. Post information was available for \num{558190} notes, corresponding to a missing rate of \SI{17.03}{\percent}.  

To construct a matched sample of request events, we iteratively retrieved posts for request entries (drawing request events at random without replacement) until we obtained the same number of request events with accessible posts as note events with accessible posts. This required attempting to fetch post data for \num{654834} requests, of which \num{558190} were accessible (missing rate of \SI{14.76}{\percent}).

Together, the \num{558190} accessible notes and \num{558190} accessible requests link to \num{711914} distinct posts, which form the basis for the analyses that follow.

\subsection{Annotation of Language, Sentiment \& Topic}

We classified the language, sentiment, and topic of the source posts using the 27\-/billion\-/parameter version of Google's open\-/weights model Gemma~3 \citep{Gemma.2025}. The model was provided with both text and images (bimodal input), and, when applicable, with quoted or replied\-/to posts as additional context. Language labels were returned as ISO codes and offer greater accuracy than the language field present in the raw X post data, as the model incorporates visual content and contextual text. 
Sentiment was categorized as \emph{Negative}, \emph{Neutral}, or \emph{Positive}. Topic classification was multi\-/label across eight dimensions: \emph{Politics and Government}, \emph{Economy and Business}, \emph{Health and Medicine}, \emph{Science and Technology}, \emph{Sports, Entertainment, and Celebrity Culture}, \emph{Social Issues and Activism}, \emph{War and Armed Conflict}, and \emph{Crime and Law Enforcement}; posts could therefore receive multiple, one, or no topic labels. Further details on this classification process are provided in the SI.

\textbf{Validation:} To validate the automated classifications, we tasked a trained research assistant (wage: \SI{15.19}{\sieuro\per\hour}) with manually annotating \num{200} randomly selected posts for language, sentiment, and topic, and compared these labels to the model's predictions. Language identification showed near\-/perfect agreement with a Macro \(F_1\) of \SI{98.76}{\percent}. Sentiment classification achieved a Macro \(F_1\) of \SI{75.77}{\percent} and, accounting for its ordinal structure, a quadratic weighted Cohen's \(\kappa\) of \SI{75.85}{\percent}, indicating substantial agreement. Topic classification also performed strongly, with a Macro \(F_1\) of \SI{96.38}{\percent} across the eight categories. Altogether, these results confirm that the LLM-based annotation approach provides reliable content characterization in our setting.

\section{Empirical Analysis}

We empirically analyze the characteristics of posts targeted by requests (demand) and those that ultimately receive fact\-/checks (supply) on X's Community Notes platform (RQ\,1). Subsequently, we use a quasi\-/experimental design to estimate the effect of displaying user requests on subsequent fact\-/check creation (RQ\,2). 

\subsection{Supply vs. Demand (RQ\,1)}

We analyze the characteristics of posts that become the targets of notes (\ie, fact\-/checking supply) and note requests (\ie, fact\-/checking demand). The unit of observation is the individual annotation rather than the post, since a single post may receive multiple notes or requests and can therefore appear multiple times in the data. In total, the analysis comprises \num{558190} notes and \num{558190} requests on a total of \num{711914} distinct posts.

\begin{figure}[t]
    \centering
    \includegraphics{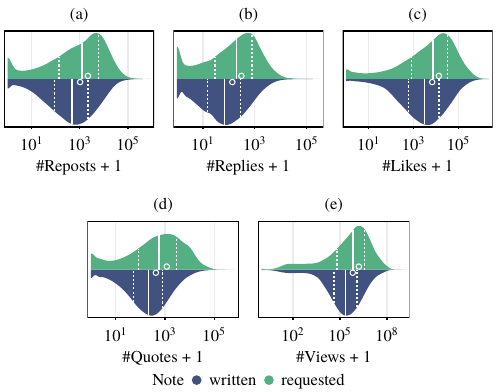}
    \caption{Split violin plots for the engagement metrics of posts targeted by note writers and requesters. Shown are kernel density estimates (areas), mean values (circles), and quartile borders (lines).}
    \label{fig:post-engage}
\end{figure}

\textbf{Engagement:} 
Across all engagement metrics (\ie, reposts, re\-plies, likes, views), requests are disproportionately directed at posts with higher visibility and interaction compared to those targeted by written notes (see Fig.~\ref{fig:post-engage}). On average, posts with requests receive \num{5284} reposts -- more than twice as many as posts with notes (\num{2646}; KS-test: $D$ = \num{0.165}; $p < \num{0.001}$). They also attract nearly three times as many replies (\num{3063} vs. \num{1111}; $D$ = \num{0.198}; $p < \num{0.001}$), almost twice as many likes (\num{32678} vs. \num{17742}; $D$ = \num{0.158}; $p < \num{0.001}$), and more than twice as many quotes (\num{734} vs. \num{360}; $D$ = \num{0.188}; $p < \num{0.001}$). Furthermore, requested posts garner over twice the number of views compared to noted posts (\num{4764041} vs. \num{2235985}; $D$ = \num{0.178}; $p < \num{0.001}$). These differences indicate that user demand is concentrated on highly visible, high\-/engagement posts, whereas contributor supply extends more broadly.

\begin{figure}[t]
    \centering
    \includegraphics{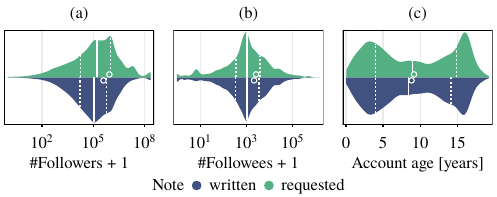}
    \caption{Split violin plots for the engagement metrics and account ages of users targeted by note writers and requesters. Shown are kernel density estimates (areas), mean values (circles), and quartile borders (lines)\\}
    \label{fig:user-engage}
    \includegraphics{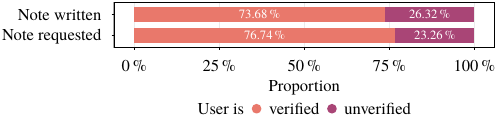}
    \caption{Distribution of the verification status among users targeted by note writers and requesters.}
    \label{fig:user-verify}
\end{figure}

\textbf{Author characteristics:}
We also observe differences in the characteristics of post authors (see Fig.~\ref{fig:user-engage}). Posts with requests originate from users averaging \num{5920461} followers -- more than twice as many as those whose posts received notes (\num{2738779}; KS-test: $D$ = \num{0.088}; $p < \num{0.001}$). Authors of requested\-/note posts also follow more accounts on average (\num{9100} vs. \num{6381}; $D$ = \num{0.028}; $p < \num{0.001}$), and their accounts are slightly older (\num{9.16} vs. \num{8.83} years; $D$ = \num{0.060}; $p < \num{0.001}$). We further observe a small but statistically significant difference in verification status: \SI{76.74}{\percent} of posts with requests come from verified users, compared to \SI{73.68}{\percent} for posts with notes ($\chi^2$-test: $X^2$ = \num{1398.719}; $p < \num{0.001}$; see Fig.~\ref{fig:user-verify}). These differences suggest that requests are more likely to target posts from prominent, well\-/connected, and longer\-/established accounts.

\begin{figure}[t]
    \includegraphics{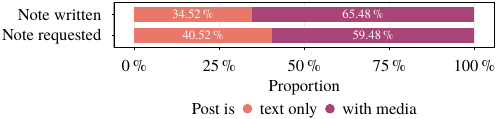}
    \caption{Distribution of the media attachment status of posts targeted by note writers and requesters.\\}
    \label{fig:post-media}
    \includegraphics{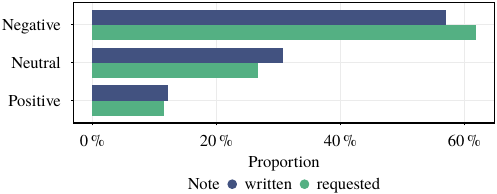}
    \caption{Distribution of the sentiment values among posts targeted by note writers and requesters.}
    \label{fig:post-senti}
\end{figure}

\textbf{Post type:}
We examine whether the posts targeted by notes and requests contain attached media. \SI{65.48}{\percent} of notes targeted posts with an attached image or video, compared to \SI{59.48}{\percent} of requests that targeted such posts, a statistically significant difference ($\chi^2$-test: $X^2$ = \num{4292.580}; $p < \num{0.001}$; see Fig.~\ref{fig:post-media}). Furthermore, we examine two structural characteristics of targeted posts: whether a post is a conversation starter or a reply, and whether it is posted with or without a quote. For replies, modest but clear differences emerge: \SI{6.61}{\percent} of notes target reply posts, compared to \SI{9.81}{\percent} of requests, a statistically significant difference ($\chi^2$-test: $X^2$ = \num{3799.471}; $p < \num{0.001}$). In contrast, the distributions for quote posts are nearly identical: \SI{12.89}{\percent} of notes and \SI{12.81}{\percent} of requests are on posts that quote another post, with no statistically significant difference ($\chi^2$-test: $X^2$ = \num{1.678}; $p = \num{0.195}$). Overall, requests are somewhat more likely than notes to focus on replies, whereas quote posts are targeted at nearly identical rates.

\textbf{Sentiment:}
We compare the sentiment of posts targeted by notes and note requests (see Fig.~\ref{fig:post-senti}). Both groups concentrate predominantly on negative content, but requests are more likely than notes to target such posts (\SI{61.82}{\percent} vs. \SI{57.09}{\percent}). For neutral sentiment, notes are more common than requests (\SI{30.73}{\percent} vs. \SI{26.66}{\percent}). For positive sentiment, the shares are relatively balanced (\SI{12.18}{\percent} vs. \SI{11.52}{\percent}). Overall, requests are somewhat more concentrated on negative content, whereas written notes more often address neutral posts. While these differences are comparatively small, they are statistically significant ($\chi^2$-test: $X^2$ = \num{2767.209}; $p$ $<$ \num{0.001}).

\begin{figure}[t]
    \includegraphics{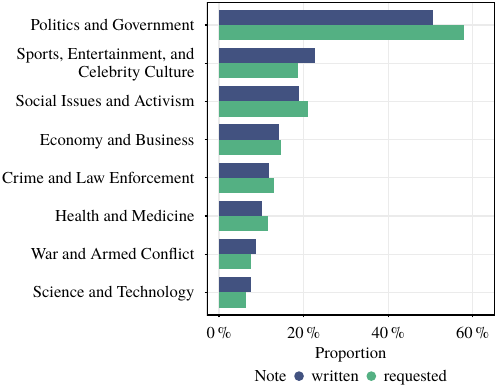}
    \caption{Distribution of the topics covered in posts targeted by note writers and requesters. Because topics are multi\-/label, aggregated proportions do not sum to \SI{100}\percent.\\}
    \label{fig:post-topic}
    \includegraphics{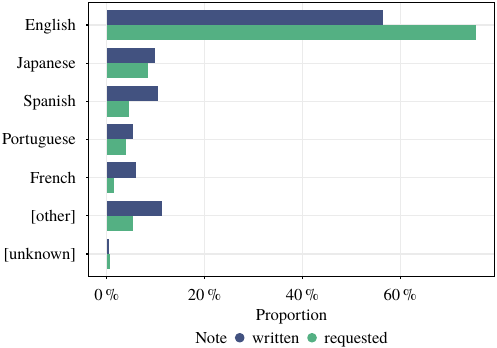}
    \caption{Distribution of the languages among posts targeted by note writers and requesters.}
    \label{fig:post-lang}
\end{figure}

\textbf{Topics:} 
We also analyze the topical distribution of targeted posts (see Fig.~\ref{fig:post-topic}). The vast majority concern \emph{Politics and Government}, with requests more concentrated than notes (\SI{57.98}{\percent} vs. \SI{50.65}{\percent}). Requests also make up slightly larger shares in \emph{Economy and Business} (\SI{14.75}{\percent} vs. \SI{14.12}{\percent}), \emph{Health and Medicine} (\SI{11.46}{\percent} vs. \SI{10.06}{\percent}), \emph{Social Issues and Activism} (\SI{21.14}{\percent} vs. \SI{18.99}{\percent}), and \emph{Crime and Law Enforcement} (\SI{13.01}{\percent} vs. \SI{11.76}{\percent}). By contrast, notes are more common for \emph{Sports, Entertainment, and Celebrity Culture} (\SI{22.67}{\percent} vs. \SI{18.73}{\percent}), \emph{Science and Technology} (\SI{7.44}{\percent} vs. \SI{6.30}{\percent}), and \emph{War and Armed Conflict} (\SI{8.77}{\percent} vs. \SI{7.67}{\percent}). Although the magnitude of these differences is modest, they are statistically significant overall ($\chi^2$-test: $X^2$ = \num{6558.860}; $p < \num{0.001}$). Taken together, this indicates that user demand is more concentrated on politics and social issues, while contributors distribute supply more evenly across a broader set of topical domains.  

\textbf{Languages:}
We further examine the language of targeted posts (see Fig.~\ref{fig:post-lang}). Both notes and requests predominantly target English posts, but requests are much more likely than notes to do so (\SI{75.56}{\percent} vs. \SI{56.48}{\percent}). For Japanese, requests are slightly less common than notes (\SI{8.39}{\percent} vs. \SI{9.91}{\percent}), while for Spanish and French the differences are more pronounced, with notes far more frequent (Spanish: \SI{10.54}{\percent} vs. \SI{4.51}{\percent}; French: \SI{6.01}{\percent} vs. \SI{1.47}{\percent}). Portuguese shows a smaller gap, again with notes more common (\SI{5.34}{\percent} vs. \SI{3.97}{\percent}). For all remaining languages (including those that could not be classified), notes are more frequent overall. Overall, requests are thus disproportionately directed at English posts, while notes are more broadly distributed across other languages. The differences in distributions are statistically significant ($\chi^2$-test: $X^2$ = \num{68108.882}; $p$ $<$ \num{0.001}). 

Taken together, these results indicate that requests tend to target posts from socially influential accounts with high levels of engagement, whereas written notes are distributed across posts with comparatively lower levels of visibility and interaction.

\subsection{Effect of Request Displays on Note Creation (RQ\,2)}

To assess whether request prompts shape contributor behavior, we leverage a quasi\-/experimental setting created by the platform's contributor hierarchy. On Community Notes, only a subset of contributors, those designated as \textit{Top Writers} based on their past history of producing highly rated notes, are permitted to view request displays. All other contributors (\ie, non\-/Top Writers) cannot see requests. This platform rule enables a within\-/post comparison of how note creation changes for Top Writers relative to non\-/Top Writers before and after a request becomes visible on the same post. In doing so, we can isolate whether requests increase the responsiveness of eligible contributors and help align the supply of fact\-/checking with user demand.

\textbf{Request display and Top Writer status:}  The public notes and requests dumps do not directly encode when requests became visible or which authors were eligible to see them. We therefore reconstruct both quantities from the released data.

At the time of data collection, a post qualified for request display if (i) it had received at least \num{4} user requests and (ii) its request rate reached one per \num{25000} views, provided the post was less than \num{24} hours old. Once a post qualified, the request prompt remained visible for \num{24} hours. Because we observe the exact time series of requests but not view counts at the time of each request, we estimate request rates under the assumption that the rate of requests per view is constant over time (see SI
for details). Under these rules, \SI{6.7}{\percent} of posts with at least one request qualified for request display to Top Writers.

Only Top Writers are eligible to see request displays. The Community Notes algorithm defines a contributor as a \emph{Top Writer} at time \(t\) if they have (i) a writing impact score of at least \num{10}, where writing impact is the difference between the number of their notes rated ``helpful'' and the number rated ``not helpful,'' and (ii) an impact rate of at least \SI{4}{\percent}, calculated as writing impact divided by their total number of published notes. Using the released note status history, we reconstruct whether each author met these criteria at the time they wrote each note (see SI
for details). Overall, \SI{19.16}{\percent} of notes were authored by Top Writers.

\begin{figure}[t]
    \centering
    \includegraphics{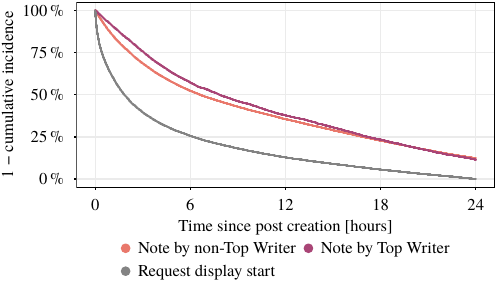}
    \caption{Kaplan\-/Meier estimates of the survival function for note creation by Top Writers and non\-/Top Writers, alongside request display onset. Time is measured relative to post creation. The y\-/axis shows 1 \(-\) cumulative incidence, meaning the proportion of contributor\-/post intervals in which the event of interest (note creation or request display) has not yet occurred at each time point.}
    \label{fig:note-surv-event}
\end{figure}

\textbf{Regression sample:}  
Our analytic sample contains all \num{19868} posts that had a request displayed and received at least one community note within the first four days after posting. These posts received a total of \num{33577} notes: \num{28345} written by non\-/Top Writers and \num{5232} written by Top Writers. For the analyses that follow, the unit of observation is at the note level. On average, request displays began \SI{4.50}{\hour} after post creation. Note creation occurred substantially later, at an average of \SI{11.4}{\hour} after posting, with similar means for non\-/Top Writers (\SI{11.3}{\hour}) and Top Writers (\SI{11.9}{\hour}); see Fig.~\ref{fig:note-surv-event}.

\textbf{Model specification:}  
To capture the timing of note creation while accounting for variation across posts, we estimate a Cox proportional hazards model with post\-/level strata \cite{Andersen.1982,Therneau.2001}. This approach allows baseline hazards to differ by post, ensuring that unobserved heterogeneity across posts does not confound the estimated effects of request visibility. The model includes both Top and non\-/Top Writers in the risk set, but only Top Writers are eligible to see request prompts. 
Formally, our empirical model is
\begin{gather*}
h_i(t)
=
h_{0,\,\mathit{post}_i}(t)\,
\exp\bigl(\eta_i(t)\bigr), \\[4pt]
\eta_i(t)
=
\beta_1\,\mathit{Top}_i
+
\sum_{k}\gamma_k\,\mathit{Top}_i
\times
\mathit{Display}_{\mathit{post}_i}^{\,k}(t), \\
\text{for all }
k \in \{[0,6)\,\si{\hour},[6,12)\,\si{\hour},[12,24]\,\si{\hour}\}.
\end{gather*}

\noindent
The term \(h_{0,\,\mathit{post}_i}(t)\) denotes a post\-/specific baseline hazard. \(\mathit{Top}_i\) is an indicator equal to 1 if the contributor of note \(i\) is a Top Writer at time \(t\), and \(\mathit{Display}_{\mathit{post}_i}^{\,k}(t)\) is a set of time\-/varying indicators that equal 1 when a request is currently displayed on the associated post, where cohorts \(k\) are defined by the post's age at the time the request is first displayed. We split the treatment into cohorts to allow the effect of request visibility to vary over the post lifecycle and to satisfy the proportional hazards assumption. The coefficient \(\beta_1\) captures the baseline difference in note creation between Top and non\-/Top Writers in the absence of a visible request. The main effect of request visibility is not included because, under post stratification, any covariate that is common to all contributors within a post at a given time is absorbed into the post\-/specific baseline hazard and is therefore not separately estimable. Identification therefore comes from the interaction terms \(\gamma_k\), which compare the behavior of Top Writers (who can see the request prompt) to non\-/Top Writers (who cannot) on the same post and at the same time.

All models were estimated in \emph{R} 4.5.1 \cite{R.core} using the \emph{survival} package \cite[version 3.8-3;][]{R.survival}.

\begin{figure}[t]
    \centering
    \includegraphics{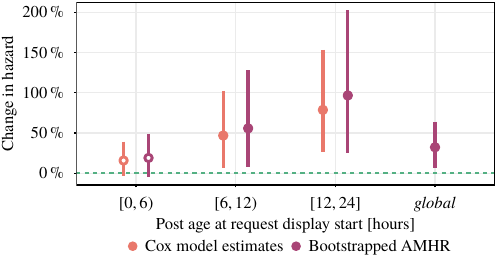}
    \caption{Impact of request displays on the hazard of note creation for Top Writers. Orange points show hazard ratios from the stratified Cox model; red points show bootstrapped Average Marginal Hazard Ratios (1,000 post\-/level replications).}
    \label{fig:placeholder}
\end{figure}

\textbf{Estimation results:}  
In the absence of a request display, Top Writers are about \SI{27}{\percent} less likely to create a note compared to non\-/Top Writers (HR = \num{0.73}; \SI{95}{\percent} CI [\num{0.62}, \num{0.86}]). When a request is displayed, the hazard of note creation increases differentially for Top Writers depending on the post's age at which the request was shown. For requests displayed within the first \SIrange{0}{6}{\hour} of a post's life, Top Writers' hazard is about \SI{16}{\percent} higher, though not statistically significant (HR = \num{1.16}; \SI{95}{\percent} CI [\num{0.96}, \num{1.39}]). Between \SIrange{6}{12}{\hour}, the effect becomes significant, with Top Writers' hazard about \SI{47}{\percent} higher compared to non\-/Top Writers (HR = \num{1.47}; \SI{95}{\percent} CI [\num{1.07}, \num{2.02}]). For requests displayed between \SIrange{12}{24}{\hour}, the effect is strongest: Top Writers' hazard of note creation nearly doubles (HR = \num{1.79}; \SI{95}{\percent} CI [\num{1.26}, \num{2.53}]). 

These results imply that displaying a request increases the likelihood that a Top Writer creates a note, and that this effect depends on when in a post's lifecycle the request is triggered. In the absence of a request, Top Writers are, on average, less likely to create notes than non\-/Top Writers, consistent with a more selective pattern of contribution. When a request prompt becomes visible, this baseline difference is reduced and, for requests triggered later in a post's lifecycle, reversed. This suggests that requests are particularly effective at directing Top Writers' attention to posts for which fact\-/checking demand remains unmet. 

To complement these coefficient estimates, we compute \emph{Average Marginal Hazard Ratios} (AMHRs). For each contributor and risk interval, we compare the predicted hazard of note creation when a request is displayed to the counterfactual hazard when no request is displayed, then average across contributor groups and intervals weighted by their empirical frequency.\footnote{Uncertainty (\ie, confidence intervals) is quantified via \num{1000} post\-/level bootstrap replications.} The global AMHR indicates that displaying requests increases the hazard of note creation by \SI{32}{\percent} on average (AMHR = \num{1.32}; 95\% CI [\num{1.06}, \num{1.64}]). Interval\-/specific AMHRs align with the coefficient estimates: modest when requests appear early (\SIrange{0}{6}{\hour}: AMHR = \num{1.20}; 95\% CI [\num{0.95}, \num{1.49}]), stronger for mid\-/aged posts (\SIrange{6}{12}{\hour}: AMHR = \num{1.59}; 95\% CI [\num{1.08}, \num{2.28}]), and largest for later requests (\SIrange{12}{24}{\hour}: AMHR = \num{2.00}; 95\% CI [\num{1.26}, \num{3.02}]).

Overall, we find that the requests feature on X substantially increases the likelihood that content receives a Community Note, indicating that it can effectively direct contributor effort toward the content that users request to be fact\-/checked.

\section{Discussion}

\textbf{Relevance:} Misinformation is a pressing challenge for social media platforms \cite{Bar.2023,Ecker.2024,WEF.2024}. Fact\-/checking is a central tool to mitigate its spread, yet fact\-/checkers have limited time and resources, making it unrealistic to verify all (or even most) content. As a result, the effectiveness of fact\-/checking systems depends not only on how many fact\-/checks are produced, but also on whether they target the content for which there is demand by users. However, prior research has typically examined either demand or supply in isolation \cite{Pilarski.2024,Renault.2025,Chuai.2025b,Rich.2020,Graham.2025,Chopra.2022}, offering little evidence on how these forces interact in real\-/world social media environments. Here, we contribute through a large\-/scale empirical analysis of supply vs. demand on X's Community Notes platform. 

\textbf{Research implications:} Our findings highlight a gap between the content users request fact\-/checks for and the content that ultimately receives fact\-/checks. Specifically, we find that requests disproportionately target highly visible posts (\ie, those with more views and engagement and authored by influential accounts), where\-as fact\-/checks are distributed more broadly across languages, sentiments, and topics. This suggests that community fact\-/checkers, as a self\-/selected group, are not fully mirroring the broader user base on social media. Such divergence has dual implications: on the one hand, it broadens the range of content that receives scrutiny; on the other, it may come at the expense of leaving the most relevant posts comparatively under\-/checked. 

More broadly, our results underscore the need for research that accounts for both demand and supply when characterizing misinformation on social media. Most of the existing empirical work (\eg, studies of diffusion, virality, and intervention effects) has focused on the supply side of fact\-/checking, \ie, analyzing posts that have already been fact\-/checked \cite{Vosoughi.2018,Prollochs.2021a,Prollochs.2022b,Prollochs.2021b,Chuai.2024b}. Yet, our results indicate that this subset may not fully represent what users want verified, introducing selection effects that could potentially bias conclusions about how misinformation spreads and which content is most consequential. Future studies may partially account for this by comparing fact\-/checked posts to demand signals (such as requests) and incorporating these signals into analyses. Doing so could potentially yield more reliable prevalence estimates and a clearer picture of the volume and characteristics of potentially misleading content that remains unverified. Such insights may help researchers identify where current fact\-/checking systems fall short and inform the design of interventions to close coverage gaps.

\textbf{Platform implications:} Our results point to opportunities for improving the design of fact\-/checking systems. We find that the requests feature on X significantly accelerates fact\-/checking activity, suggesting that it can effectively steer contributor effort toward content users want to be fact\-/checked. In the future, platforms could build on this by systematically monitoring posts that are particularly concerning but remain unchecked and treating such demand\-/supply gaps as a governance signal. Such monitoring could guide the targeted routing of requests to contributors with relevant expertise (\eg, by topic, language) and help allocate scarce fact\-/checking resources to the most concerning misleading posts. Moreover, platforms could treat demand\-/supply alignment as an explicit performance metric, \ie, using it to detect persistent coverage gaps (\eg, underserved languages or low\-/visibility but harmful content) and to rigorously evaluate the effect of platform interventions on closing those gaps.

\textbf{Limitations and future research:} As with any other research, our study is not free of limitations. First, our analysis is limited to a single platform, \ie, community\-/based fact\-/checking on X. Future research could explore whether similar patterns hold across other platforms or cultural settings. Second, requests capture only one type of demand signal and may underrepresent the concerns of users who are ineligible or unaware of the feature. Combining requests with complementary signals such as user reports or survey data could yield a more complete picture of demand. Third, the underlying data impose constraints, as a portion of community\-/noted posts is no longer accessible via the API due to deletion or account suspensions. Relatedly, we had to reconstruct the display status and timing of request callouts because these values are not included in the published Community Notes data and are also not available through the X API's post endpoint. 
Greater transparency from the platform, for example, by publishing request display logs or historical view metrics, may enhance the precision of future research in this area. Fourth, an important open question concerns the motivations of requesters: future work could examine whether requests primarily reflect genuine demand for verification or are sometimes driven by political or coordinated behavior. Fifth, incorporating information on political affiliation could yield additional insights. For example, recent research indicates that fact\-/checking contributors more frequently annotate posts from Republicans, whereas requests disproportionately target Democratic posts~\cite{Chuai.2025c}. 
Sixth, our findings highlight that user demand tends to concentrate on highly visible content, but visibility is not always synonymous with harm. Future work should disentangle attention (what is seen and requested) from potential harm (what is likely to misinform or cause real\-/world impact). Finally, while our quasi\-/experimental design provides evidence that requests influence fact\-/checking activity, further research could examine heterogeneous treatment effects (\eg, across languages, topics, or contributor experience levels) and potential trade\-/offs, such as whether prioritizing requested posts crowds out coverage of other content.

\section{Conclusion}

This paper provides the first large\-/scale analysis of supply and demand in community\-/based fact\-/checking. Using data from 1.1 million notes and requests on X's Community Notes platform, we show that demand tends to concentrate on highly visible posts from influential accounts, while supply is distributed more broadly across languages, sentiments, and topics. We further find that displaying requests significantly accelerates fact\-/checking contributions, suggesting that request mechanisms can help align contributor effort with user priorities. Together, these findings highlight both structural imbalances in community\-/based fact\-/checking and the potential of design interventions to mitigate them.

\section{Ethics Statement}

All analyses are based on publicly available data. The data collection and the analysis follow common standards for ethical research \cite{Rivers.2014}. We declare no competing interests.

\bibliography{references.bib}

\clearpage
\appendix

\section*{Supplementary Information}\label{sec:si}

\subsection{Request Display and Top Writer Status}\label{sec:si-rqst-top}

The public notes and requests dumps do not directly encode when requests became visible or which authors were eligible to see them. We therefore reconstruct both quantities from the released data.

\textbf{Request display status.}  
A request display for a post is visible to Top Writers at time \(t\) if 
(i) it has received at least \num{4} user requests, (ii) its request rate is at least one per \num{25000} views, and (iii) its age is less than \SI{24}{\hour}.
Once visible, the request prompt remains displayed to Top Writers for \SI{24}{\hour}. 

Because view counts at the time of each request are not directly reported, we estimate views for each post \(p\) at time \(t\) -- corresponding to the moment a request is submitted -- under the assumption that requests arrive in proportion to total views:
\[
\mathit{CurrentViews}_{p,t} \approx \mathit{TotalViews}_p \times \frac{\mathit{CurrentRequests}_{p,t}}{\mathit{TotalRequests}_p}.
\]

\begin{figure}[h!]
    \centering
    \includegraphics{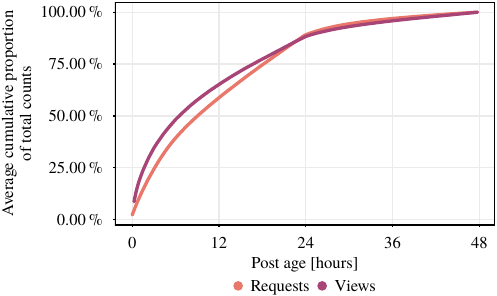}
    \caption{Average cumulative proportion of total request and view counts as a function of post age. Request counts are computed for posts in our regression sample; view counts are taken from the dataset published alongside the paper by \citet{Slaughter.2025}.}
    \label{fig:post_rqst_view_count_ts}
\end{figure}

To assess the plausibility of this assumption, we compare the request count time series for all posts in our regression sample to the view count time series from the reproducibility dataset published alongside the paper by \citet{Slaughter.2025}, which contains posts that received Community Notes and provides detailed view trajectories. Figure~\ref{fig:post_rqst_view_count_ts} plots the average cumulative proportions of total requests and total views as a function of post age. The two curves follow a broadly similar trajectory, suggesting that requests and views accumulate on comparable time scales and supporting our proportional\-/arrival approximation.

\noindent
We then infer the start of the request display interval for each post \(p\) at time \(t\), corresponding to the moment a request is submitted, as:
\[
\begin{aligned}
\mathit{RequestDisplay}_{p,t} =\: & 
\mathit{CurrentRequests}_{p,t} \ge \num{4} \\
\text{and}\quad &
\frac{\mathit{CurrentRequests}_{p,t}}{\mathit{CurrentViews}_{p,t}} \ge \frac{1}{\num{25000}} \\
\text{and}\quad &
\mathit{PostAge}_{p,t} < \SI{24}{\hour}.
\end{aligned}
\]

\textbf{Top Writer status.}  
Only Top Writers are eligible to see request displays. Top Writer status depends on both the volume and proportion of a contributor's notes that are rated as helpful by others. Formally, Top Writer status for contributor \(c\) at time \(t\) is defined in the open\-/source Community Notes algorithm \citep{CoNo.Code.TopWriter} as:
\[
\begin{aligned}
\mathit{WritingImpact}_{c,t} =\: & 
\mathit{HelpfulNotes}_{c,t} \\ & - \mathit{NotHelpfulNotes}_{c,t}, \\
\mathit{IsTopWriter}_{c,t} =\: & 
\mathit{WritingImpact}_{c,t} \geq 10 \\
\text{and} \quad & 
\frac{\mathit{WritingImpact}_{c,t}}{\mathit{TotalNotes}_{c,t}} \geq 0.04.
\end{aligned}
\]
We compute these quantities dynamically and evaluate Top Writer status immediately prior to the creation of each note, allowing us to determine whether the contributor was eligible to see request displays at the time the note was published.

\subsection{LLM-Based Annotation}\label{sec:si-llm}

To annotate the content of posts targeted by note writers and requesters with respect to their language, sentiment, and topics, we use Google's open\-/weights model Gemma~3 \citep{Gemma.2025} in the 27\-/billion\-/parameter configuration with 4\-/bit quantization. The model was served using OLLAMA (version~0.9.1).  

The model was provided with both text and images (bimodal input), and, when applicable, with quoted or replied\-/to posts as additional context.  

We requested structured output by supplying a template JSON schema via the OLLAMA model API and by including corresponding instructions in the system prompt. Language and sentiment were returned as enumerated fields (from a predefined set of possible values), while topic annotations were returned as a JSON object containing multiple boolean fields (one per topic dimension). An example of the schema specification is included in the system prompt below.

\subsection{System Prompt}

{
\ttfamily
\raggedright\sloppy
\setlength{\parindent}{0pt}
\setlength{\parskip}{\baselineskip}
\footnotesize
\noindent
You are a research assistant tasked with classifying social media posts based on their language, sentiment, and topical content. Each post may contain text, an image, or both. Images may include video previews.

\#\# Input Structure

Each input consists of three sections:\\
- **Main Post to be Classified**\\
- **Replied\-/To Post**\\
- **Quoted Post**

The **Main Post to be Classified** is always present. If the post is not a reply or does not quote another post, the corresponding sections will contain the placeholder `*[None]*`.

Use the *Replied\-/To Post* and *Quoted Post* solely as context. Your classification must be based **exclusively** on the *Main Post to be Classified*.

\#\# Output Structure

Respond with a JSON object in the following structure — **no extra text, formatting, or explanation**:

```json\\
\{\\
\ \ "language": "en",\\
\ \ "sentiment": "Neutral",\\
\ \ "topics": \{\\
\ \ \ \ "Politics and Government": false,\\
\ \ \ \ "Economy and Business": false,\\
\ \ \ \ "Health and Medicine": false,\\
\ \ \ \ "Science and Technology": false,\\
\ \ \ \ "Sports, Entertainment, and Celebrity Culture": false,\\
\ \ \ \ "Social Issues and Activism": false,\\
\ \ \ \ "War and Armed Conflict": false,\\
\ \ \ \ "Crime and Law Enforcement": false\\
\ \ \}\\
\}\\
```

\#\# Labeling Instructions

\#\#\# `"language"`

Identify the language of the *Main Post*. Return the ISO 639-1 code (e.g., `"en"` for English, `"de"` for German). If the language is unclear or unrecognizable, return `"[unknown]"`.

\#\#\# `"sentiment"`

Determine the overall sentiment of the *Main Post*. Select exactly one of the following values: `"Negative"`, `"Neutral"`, or `"Positive"`.

\#\#\# `"topics"`

Identify the main topics clearly discussed in the Main Post. Return an object where each predefined topic key has a boolean value. Set each value to `true` only if the topic is a clear and substantial focus of the post; otherwise, set to `false`. Avoid tagging topics that are only referenced in passing or implied indirectly. Emphasize precision over coverage.

\#\#\#\# Valid topic keys

- `"Politics and Government"` — e.g., domestic and foreign policy, elections, legislation, public officials\\
- `"Economy and Business"` — e.g., markets, finance, jobs, inflation, corporate news\\
- `"Health and Medicine"` — e.g., public health, diseases, wellness, healthcare\\
- `"Science and Technology"` — e.g., research, innovation, artificial intelligence, tech news\\
- `"Sports, Entertainment, and Celebrity Culture"` — e.g., celebrities, sports events, media, fandom\\
- `"Social Issues and Activism"` — e.g., civil rights, protests, cultural debates, race, gender, LGBTQ+ issues\\
- `"War and Armed Conflict"` — e.g., military actions, invasions, battlefield reporting, armed group activity\\
- `"Crime and Law Enforcement"` — e.g., criminal events, policing, legal controversies

---

Return **only** the JSON object. No additional output or formatting.

}

\end{document}